\documentstyle[multicol,prl,aps,epsf]{revtex}
\begin{document}
\title{Cuprate core level line shapes for different Cu-O networks}    
\author{K. Karlsson$^{1,2}$, O. Gunnarsson$^2$ and O. Jepsen$^2$}  
\address{${}^{(1)}$Inst. f\"or Naturvetenskap, H\"ogskolan  Sk\"ovde,
S-541 28 Sk\"ovde, Sweden}
\address{${}^{(2)}$Max-Planck-Institut f\"ur Festk\"orperforschung, 
D-70506 Stuttgart, Germany}

\date{\today}
\maketitle
\begin{abstract}
We have studied the Cu core level photoemission spectra in 
the Anderson impurity model for cuprates with different 
Cu-O networks,    dimensionalities (zero, one, 
two and three) and  Cu valencies (two and three).  
We focus on the shape of the leading peak and obtain  
very good agreement with the experimental data. We show 
how the shape of the spectrum is related to the valence 
electronic structure and the Cu-O network but also that 
other atoms can play a role.
\end{abstract}
%\pacs{79.60.-i, 71.20.Be, 71.10.Fd}
\begin{multicols}{2}
B\"oske {\it et al.}\cite{Dresden} have shown 
that for a number of Cu compounds the shape of the {\it core}
level photoemission spectrum strongly depends on the type of 
Cu-O network. This illustrates that it is not sufficient to think 
of the core level spectra as primarily atomic-like, but that 
there is a substantial influence from the valence electron 
structure. The core level spectrum can therefore give interesting
information about the electronic structure. 

The Cu compounds studied by B\"oske {\it et al.} are divalent,
with Cu having mainly    d$^9$ character. When a Cu core hole 
is created, it is energetically favorable to transfer an 
electron from the surrounding to the site of the core hole, 
forming a d$^{10}$ state. This leads to a spectrum with a 
leading peak of primarily d$^{10}$ character and a satellite 
of mainly d$^9$ character\cite{Kotani}. The main peak 
corresponds to a process where an electron has hopped from 
the valence band to the site with a core hole, and its shape 
reflects the valence band structure\cite{CO,Krister}. This 
leads to a broadening of the main peak and possibly to further 
structures.  Early experiments with a moderate resolution 
showed little structure. The high-resolution experiments 
of B\"oske {\it et al.}\cite{Dresden}, however, reveal a 
clear system dependence of the core level spectra.      
In the present paper we calculate these spectra and show 
that they are in good agreement with experiment. We explain 
in detail the origin of the differences between different 
cuprates. It has been suggested that the shape of the leading 
peak is explained in terms of a two-peak structure, due to
so-called local and nonlocal screening\cite{Sawatzky}.
The recent experiments, however, show that the shape 
can be substantially more complicated.   

To calculate the core spectrum, we use the Anderson impurity model
\begin{eqnarray}\label{eq:1}
H&&=\sum_{\nu=1}^{10} \lbrace \int \varepsilon 
\psi^{\dagger}_{\varepsilon\nu} \psi_{\varepsilon\nu}
d\varepsilon +\lbrack \varepsilon_d -U_{dc}(1-n_c)\rbrack
\psi^{\dagger}_{\nu}\psi_{\nu}  \nonumber \\
&&+\int \lbrack V_{\nu}(\varepsilon)\psi^{\dagger}_{\nu}
\psi_{\varepsilon\nu}+h.c.\rbrack d\varepsilon\rbrace+
U\sum_{\nu<\mu}n_{\nu}n_{\mu},
\end{eqnarray} 
where $\nu$ is a combined spin and orbital index. The 
(host) valence states are labelled by the energy 
$\varepsilon$ and $\nu$, and the Cu impurity $3d$
states are labelled by $\nu$\cite{Handbook}.
The impurity level is pulled down from $\varepsilon_d$ to 
$\varepsilon_d-U_{dc}$ when a core hole is created. The 
electrons on the impurity have the Coulomb repulsion $U$.

It is crucial to use realistic hopping matrix elements
$V_{\nu}(\varepsilon)$. 
The local density of states (LDOS) $\rho_{\nu}(\varepsilon)$ 
of a $3d$ orbital with the label $\nu$ is calculated 
for all compounds studied here, using the 
the local density approximation (LDA) and the LMTO
method\cite{LMTO}. Then\cite{v2}
\begin{equation}\label{eq:2}
\pi|V_{\nu}(\varepsilon)|^2=-{\rm Im} \lbrace \lbrack \int 
d \varepsilon^{'}{\rho_{\nu}(\varepsilon^{'})\over 
\varepsilon-\varepsilon^{'}-i0}\rbrack ^{-1}\rbrace.
\end{equation}
An Anderson impurity calculation using these matrix elements 
contains effects of all the $3d$ orbitals, not just the ones  
on the impurity. An {\it impurity} calculation with 
$U=0$ exactly reproduces the LDOS of a {\it lattice} band 
structure calculation\cite{v2}. In our calculation
with $U\ne 0$, the many-body effects are treated explicitly 
on the site where the core hole is created, while it is 
assumed that the $3d$ states on the other sites can be 
treated in the LDA. In the discussion of the Anderson 
impurity model, it is often overlooked that all $3d$ states
are (approximately) included.  

We set $U=8$ eV and $U_{dc}=10$ eV\cite{Krister},
and adjust $\varepsilon_d$ so that the relative weights of 
the main peak and the satellite agree with experiment.
We have neglected multiplet effects. These
effects are important for the shape of the satellite, but 
not for the shape of the main peak, which we consider here.

To solve the model, we expand the ground-state as\cite{PR83}
\begin{equation}\label{eq:3}
|E_0\rangle=A(\psi_{\nu}|d^{10}\rangle+
\int^{\varepsilon_F} d\varepsilon a(\varepsilon)\psi_{\varepsilon\nu}
|d^{10}\rangle),
\end{equation}
where the parameters $A$ and $a(\varepsilon)$ are  
determined variationally. $|d^{10}\rangle$ is a state with 
all the valence states ($|\varepsilon \nu\rangle$) up to 
the Fermi energy ($\varepsilon_F$) as well as all the impurity 
states filled. The first term in (\ref{eq:3}) corresponds 
to a $d^{9}$ configuration and the second to a state where 
a valence electron with the energy $\varepsilon$ has filled 
the $3d$-hole, in the following referred to as 
$d^{10}\varepsilon^{-1}$.  Then          
\begin{equation}\label{eq:4}
a_{\nu}(\varepsilon)={V_{\nu}(\varepsilon)\over
\Delta E -\varepsilon_d+\varepsilon}.
\end{equation}
$\Delta E$ is the lowering of the total energy due to hopping 
relative to the energy of the $d^9$ state. Typically 
$|\Delta E-\varepsilon_d|$ is of the order of a few eV. 
A lowering of $\varepsilon$ leads to a decrease of        
$a(\varepsilon)^2$, unless the variation of the denominator
is compensated by a strong energy dependence of 
$|V(\varepsilon)|^2$.
By adding more states to the Ansatz (\ref{eq:3}), the model 
can be solved increasingly accurately. However, the effect 
of additional states is essentially just a renormalization 
of $\varepsilon_d$, which we treat as an adjustable parameter 
anyhow.  The core level spectrum is calculated, by using a resolvent 
operator and by inserting intermediate states similar to       
the two terms in Eq. (\ref{eq:3})\cite{PR83,Handbook}.
The spectra were broadened by a Lorentzian (FWHM=1.1 eV) 
to simulate life-time effects and by a Gaussian (FWHM=0.4 eV) 
to simulate the instrumental broadening. 

\begin{figure}
{\epsfxsize= 7.00cm \epsffile{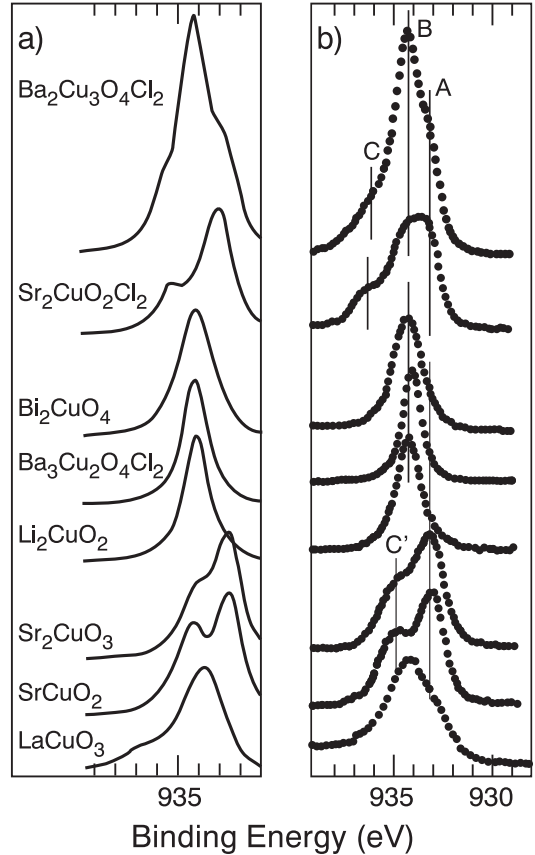}}
\caption[]{a) Theoretical core level spectra for different 
divalent and one trivalent (LaCuO$_3$) cuprate. The figure 
only shows the main peak, while the satellite at higher 
binding energies is not shown. b) Experimental spectra 
by B\"oske {\it et al.}\cite{Dresden} for the divalent 
Cu compounds and by Mizokawa {\it et al.}\cite{LaCuO3}
for the trivalent LaCuO$_3$.}
\label{fig:spectra}
\end{figure}

The results are shown in Fig. \ref{fig:spectra}, and they are  
in very good agreement with the experiments of B\"oske 
{\it et al.}\cite{Dresden}. In agreement with experiment,             
Ba$_3$Cu$_2$O$_4$Cl$_2$ and Li$_2$CuO$_2$ have very narrow peaks, 
Bi$_2$CuO$_4$ a slightly broader peak and the remaining compounds
significantly broader spectra. The shoulder of SrCuO$_2$
is more pronounced than for Sr$_2$CuO$_3$, as it should, and 
the three peak structure of Ba$_2$Cu$_3$O$_4$Cl$_2$ is 
reproduced very well.  The main difference to experiment 
is that the leading structure of Sr$_2$CuO$_2$Cl$_2$ is 
narrower. The high binding energy structures of Sr$_2$CuO$_3$ 
and SrCuO$_2$ furthermore occur at somewhat lower binding 
energy and the high binding energy structure of SrCuO$_2$ 
is somewhat more pronounced. 

\begin{figure}
{\epsfxsize= 7.50cm \epsffile{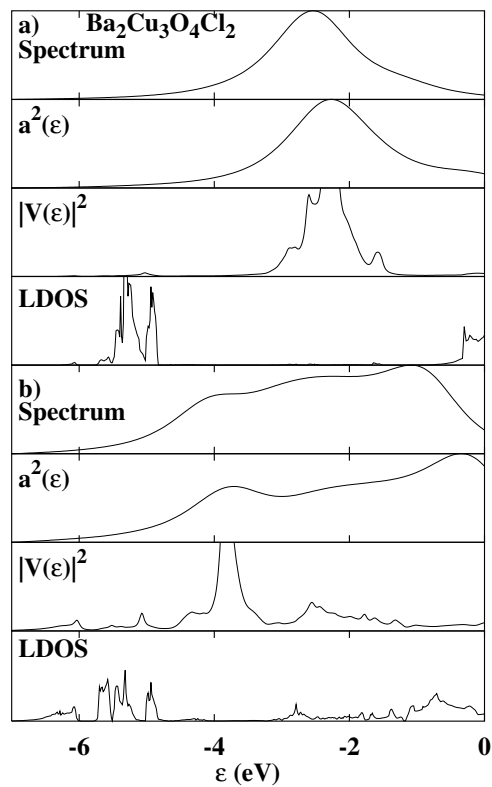}}
\caption[]{Core level spectrum, broadened $a^2(\varepsilon)$,
$|V(\varepsilon)|^2$ and local density of states (LDOS)   
for the two ( a) and b)) inequivalent Cu atoms in  
Ba$_2$Cu$_3$O$_4$Cl$_2$. $|V(\varepsilon)|^2$ and the LDOS have 
been given a small (FWHM=0.03 eV) broadening, while $a^2(\varepsilon)$
has been given the same broadening as the spectrum. We have put
$\varepsilon_F=0$.}
\label{fig:analysis}
\end{figure}

We now discuss the shape of the spectra. In the presence 
of a core hole, the $d^{10}\varepsilon^{-1}$ configuration 
is significantly lower in energy than the $d^9$ configuration. 
For the qualitative discussion (but not in the calculations) 
we assume that the coupling between these  configurations 
can be neglected in the final state. Although this 
underestimates the weight of the leading peak, it reproduces 
its shape quite well (compare `Spectrum' and `$a^2(\varepsilon)$'
in Fig. \ref{fig:analysis}). The corresponding spectrum is 
\begin{equation}\label{eq:5} 
\rho(\varepsilon)=A^2\lbrack a^2(\varepsilon)\Theta(-\varepsilon)
+\delta(\varepsilon-\varepsilon_d+U_{dc})\rbrack,
\end{equation}
where the first term gives the main peak and the second 
the satellite. The shape of the main peak is then 
determined by $a^2(\varepsilon)$, which is related to 
$|V(\varepsilon)|^2$ via Eq. (\ref{eq:4}). We therefore 
first study the relation between $V(\varepsilon)$ and 
the electronic structure.

$V(\varepsilon)$ gives the hopping between the impurity 
and host (the rest of the solid but with the impurity 
removed). We first consider the case when the LDOS 
$\rho_{\nu}(\varepsilon)$ in Eq. (\ref{eq:2}) is given 
by a semi-ellipse
\begin{equation}\label{eq:6}
\rho_{\nu}(\varepsilon)={2\over \pi B^2}\sqrt{B^2-\varepsilon^2} \hskip 1cm 
|\varepsilon| \le B,
\end{equation}
where $2B$ is the width of the band.
Performing the Hilbert transform in Eq. (\ref{eq:2}) then gives
the same shape
\begin{equation}\label{eq:7}
\pi|V_{\nu}(\varepsilon)|^2={1\over 2\pi} \sqrt{B^2-\varepsilon^2}
={B^2\over 4}\rho_{\nu}(\varepsilon) \hskip 1cm 
|\varepsilon| \le B.
\end{equation}
The overall strength of the hopping is determined by 
the width of the LDOS. This is expected, since a large 
width of the LDOS implies a strong hopping.  

The most important energy bands in the cuprates studied
here consist of antibonding combinations of one Cu orbital
(e.g., $x^2-y^2$) with $p$ orbitals on the neighboring
O sites located around $\varepsilon_F$ and corresponding
bonding combinations some 5-6 eV below $\varepsilon_F$.
The bonding-antibonding splitting is 
mainly determined by the Cu-O hopping, while the width 
of the two bands mainly depends on the hopping to 
more distant CuO$_n$ units. As a model for the LDOS we use 
two semi-ellipses  

\begin{eqnarray}\label{eq:8} 
&&\rho_{\nu}(\varepsilon)=\alpha{2\over \pi B^2} \sqrt{B^2-\varepsilon^2}  
\Theta(B^2-\varepsilon^2) \\ &&+(1-\alpha){2\over \pi B^2} \sqrt{B^2-
(\varepsilon-\varepsilon_0)^2}\Theta(B^2-(\varepsilon-\varepsilon_0)^2).
\nonumber
\end{eqnarray}
This LDOS may result in a $\delta$-function between the two 
continuous parts of the LDOS. If $\alpha|\varepsilon_0|>>B$ and 
$(1-\alpha)|\varepsilon_0|>>B$, there is a $\delta$-function
at $\varepsilon=\alpha\varepsilon_0$ with the weight $\pi\alpha
(1-\alpha)\varepsilon_0^2$. This corresponds to the 
coupling to a linear combination of $p$ orbitals on the neighboring 
O atoms. The position of the  $\delta$-function depends, however,
on the weights and widths of the bonding- and antibonding bands.
In addition, there is a continuous part at the bonding and
antibonding bands with a strength that is related to the widths
of these bands. In practice, the LDOS is not zero between the bonding-
and antibonding bands, and the $\delta$-function then has a           
broadening. 

In Fig. \ref{fig:analysis} we show results for 
Ba$_2$Cu$_3$O$_4$Cl$_2$, which has two inequivalent Cu atoms
(Cu$_A$  and Cu$_B$). One of the Cu atoms (Cu$_B$) is 
connected to neighboring Cu$_B$ atoms only via 
Cu-O-O-Cu bond and to neighboring Cu$_A$ atoms
via 90$^{\circ}$ Cu-O-Cu bond angels. Neighboring  Cu 
atoms then couple to different O $p$ orbitals, and they 
only have a weak coupling via other orbitals. The bonding- 
and antibonding bands are therefore  narrow, as illustrated 
by the panel LDOS in Fig. \ref{fig:analysis}a. The resulting     
$|V(\varepsilon)|^2$  has a broadened $\delta$-function 
between the bands but very little weight at the energies 
of the bands.  The corresponding $a^2(\varepsilon)$  
also has most of the weight between the two bands. This shape 
agrees well with the result of the full calculation (without 
the approximation (\ref{eq:5})), and explains the narrow peak 
in the core level spectrum from Cu$_B$. The other Cu 
atom (Cu$_A$) is connected to the neighboring Cu atoms via 
180$^{\circ}$ Cu-O-Cu bond angels. Neighboring Cu atoms 
then couple to the same O $p$ orbital, i.e., they have 
a strong coupling. As a result, in particular the 
antibonding band is broad, as illustrated by the LDOS 
in Fig. \ref{fig:analysis}b. The corresponding 
$|V(\varepsilon)|^2$ has a broadened $\delta$-function,
which is shifted towards lower energies than for Cu$_B$, 
due to the different widths and weights of the two bands. 
In particular, however, there is a substantial coupling 
in the energy range of the antibonding band. This coupling 
is strongly enhanced by the energy denominator 
($\Delta E-\varepsilon_d+\varepsilon)$ of Eq. (\ref{eq:4}) 
in the calculation of $a(\varepsilon)$.  Therefore, this 
coefficient is rather large for small binding energies. 
Due to a structure in $|V(\varepsilon)|^2$ at $\varepsilon
\sim -2$ eV, $a^2(\varepsilon)$ also has a structure at 
this energy. Finally, the large structure in $|V(\varepsilon)|^2$
at $\varepsilon\sim -4$ eV leads to a third structure in the 
spectrum. Adding the spectra from Cu$_A$ and Cu$_B$ gives 
the spectrum in Fig.  \ref{fig:spectra}.

Following the analysis in Fig. \ref{fig:analysis}, we
divide the systems in roughly two classes. One class consist
of systems where the CuO$_4$ units have a weak coupling to 
the other CuO$_4$ units,  either due to the Cu-O-Cu bond 
angels being (rather close to) 90$^{\circ}$ 
or because the systems is approximately  zero-dimensional (0d). 
In the second class the CuO$_4$ units have a strong coupling
to other CuO$_4$ units, due to Cu-O-Cu bond angels of the order
of 180$^{\circ}$.

Bi$_2$CuO$_4$ (0d), Ba$_3$Cu$_2$O$_4$Cl$_2$ and Li$_2$CuO$_2$ 
belong to the first class. We would then expect a narrow peak 
in the core spectrum, occurring at a higher binding energy than 
the leading edge. This is confirmed for the latter two systems, 
while both theory and experiment shows a somewhat broader peak for 
Bi$_2$CuO$_4$.  The reason for this is that there is a coupling between 
the CuO$_4$ units via the Bi atoms. As a result the structure in 
$|V(\varepsilon)|^2$ between the bonding- and antibonding bands
is broader, leading     to a somewhat broader core level spectrum.
This illustrates that not only the Cu-O network plays a role.

Sr$_2$CuO$_2$Cl$_2$, Sr$_2$CuO$_3$ and SrCuO$_2$ belong to the 
second class. In the two-dimensional (2d) structures 
Ba$_2$Cu$_3$O$_4$Cl$_2$ (Cu$_A$) and Sr$_2$CuO$_2$Cl$_2$
the antibonding band is broader than in the 1d structures 
Sr$_2$CuO$_3$ and SrCuO$_2$, due to the more efficient 
hopping to neighboring CuO$_4$ units in the 2d case. Therefore,
the structure in $|V(\varepsilon)|^2$ at $\varepsilon\sim -2$ eV 
is broader and the peak in $|V(\varepsilon)|^2$ at 
$\varepsilon\sim  -4$ eV has been pushed to somewhat lower 
energies in the 2d case.  The large binding energy structure 
in the main peak is then pushed towards somewhat larger energies 
in the 2d case, in agreement with experiment (see 
C and C$^{'}$ in Fig. \ref{fig:spectra}).

Above we have considered divalent Cu. We have earlier 
studied the trivalent NaCuO$_2$, and found that this 
has a narrow leading peak due to a narrow structure 
in $|V(\varepsilon)|^2$\cite{Krister}. An alternative 
explanation was suggested by a cluster calculation, 
where it was argued that the reason for the small 
width is the valence (3+)\cite{Sawatzky}. We have 
therefore considered a different trivalent compound, 
namely LaCuO$_3$. The corresponding spectrum is shown 
at the bottom of Fig. \ref{fig:spectra}. The main peak 
is broad, both according to theory and 
experiment\cite{LaCuO3}. The reason is similar as for 
the broad peaks in the divalent compounds, namely a broad 
antibonding band due to a Cu-O-Cu bonding angle of about 
180$^{\circ}$. Due to the three-dimensional character 
of this compound, the band is unusually broad, but only a 
relatively small fraction is occupied, because the compound 
is trivalent. Therefore the broadening of the leading 
peak is only modest. If we had instead studied a model of the 
type used in the cluster calculations, with only one orbital 
per atom, the antibonding band would have been empty. The 
Anderson impurity model would then have predicted a narrow 
leading peak, as the cluster calculations do. To obtain the 
correct description of LaCuO$_3$ it is therefore crucial to 
use a realistic model which includes the partial filling of the 
antibonding band, and which goes beyond the normally used cluster 
models.

In an exact diagonalization treatment of
a Cu$_3$O$_{10}$ cluster, the leading peak consists
of two $\delta$-functions. These correspond to so-called
local and nonlocal screening\cite{Sawatzky}, indicating that
the charge screening of the core hole comes from the closest
O$_4$ unit or from the neighboring CuO$_4$ units, respectively.
Similar results are obtained in the Anderson impurity model
for such a cluster. Finding the solutions of the host,       
we notice that the highest occupied level has a strong 
antibonding character and is located on the neighboring 
CuO$_4$ units. The second highest occupied level (with a 
coupling to the impurity) has a large weight on the O$_4$ 
unit around the impurity. Due to the missing hybridization 
with the impurity, this host level is less antibonding and 
lower in energy. The two $\delta$-functions in the spectrum 
essentially correspond to an electron from one of these two 
host levels hopping into the impurity orbital. Due to their 
different characters, these levels give local or nonlocal 
screening. 

It has been emphasized that it is important to take into
account the formation of a Zhang-Rice singlet on the CuO$_4$ 
units neighboring the unit with a core hole\cite{Sawatzky}. 
We find  that in the Anderson impurity model of a Cu$_3$O$_{10}$ 
cluster, the two lowest final states contain comparable amounts
of a Zhang-Rice singlet as is found in exact diagonalization
for such a cluster. If the highest occupied host level is 
emptied, a large amount of antibonding character is removed 
from the neighboring CuO$_4$ units. This alone is enough 
to give a large overlap to a state with a Zhang-Rice singlet 
on one of these  units. If instead the second highest host 
level is emptied, the screening charge is primarily taken 
from the local O$_4$ unit, and the local singlet character 
on the neighboring CuO$_4$ units remains small.  

To further test the role of the singlet
in core level photoemission, we have performed a calculation for a 
Cu$_3$O$_{10}$ cluster with 25 valence electrons. This system has
only one valence hole, and a Zhang-Rice singlet cannot form. 
The change of the number of electrons gives a somewhat different 
spectrum, but the main peak still has a substantial broadening,
illustrating that just going beyond a single site model (CuO$_4$)
leads to a broadening of the leading peak. The considerations above
illustrate the usefulness of a  picture where an electron     
from a host level hopps into the impurity orbital.

It has been found that even rather `large' clusters like
Cu$_3$O$_{10}$ may not be converged with respect to their
size\cite{Okada}. The inclusion of additional orbitals like 
O $p_{\pi}$ may also be important\cite{Okada}. Here we have 
found further examples of a need for more realistic models, e.g.,
for Bi$_2$CuO$_4$ and LaCuO$_3$. In an exact diagonalization 
approach, however, it is not possible to make the cluster
much larger or to add many other orbitals, since the many-electron
space then becomes too large, while such a system can be
treated in the Anderson impurity approach.

To summarize, we have calculated core level spectra for a number
of cuprates in the Anderson impurity model using {\it ab 
initio} hopping matrix elements. In this way, many-body effects 
are treated explicitly for the Cu site where the core hole  
is created, while the other Cu sites are treated in the LDA. 
This model gives a very good agreement with experimental core level 
spectra. The shape of the leading peak is influenced 
substantially by the valence electronic structure and thereby
the lattice structure. The presence of (approximately) 
180$^{\circ}$ Cu-O-Cu bond angels leads to a broad and structured 
leading peak, and an increase in the dimension of the Cu-O 
network tends to increase the width.  In particular for 
the trivalent compounds, it is important to go beyond the 
simple cluster models with one orbital per site, since these
models always give a narrow leading peak.

This work has been supported by the Max-Planck-Forschungspreis.

\end{multicols}
\end{document}